\documentclass[aps, prl, 10pt, twocolumn,superscriptaddress,longbibliograph]{revtex4-2}
\usepackage{
    physics,
    mathtools,
    amssymb,
    bm,
    xcolor,
    graphicx,
}
\usepackage[bookmarks=false,linkcolor=blue,urlcolor=blue,colorlinks,citecolor=blue]{hyperref}
\usepackage[capitalize]{cleveref}
\Crefname{figure}{Fig.}{Figs.}
\usepackage[utf8]{inputenc}
\usepackage{epsfig}
\usepackage{graphicx}
\usepackage{float}
\usepackage{dcolumn}
\usepackage{xcolor}
\usepackage{bm}
\usepackage[normalem]{ulem}
\usepackage{amsmath,bbm}

\DeclareUnicodeCharacter{2212}{-}

\begin{document}

\title{Gate tunable enhancement of supercurrent in hybrid planar Josephson junctions}

\author{Peng Yu}
\affiliation{Center for Quantum Information Physics, New York University, New York, NY 10003, USA}
\author{Han Fu}
\affiliation{Department of Physics, William \& Mary, Williamsburg, Virginia 23187, USA} 
\affiliation{Department of Physics, Florida Atlantic University, Boca Raton, Florida, 33431, USA}
\author{William F. Schiela}
\affiliation{Center for Quantum Information Physics, New York University, New York, NY 10003, USA}
\author{William Strickland}
\affiliation{Center for Quantum Information Physics, New York University, New York, NY 10003, USA}
\author{Bassel Heiba Elfeky}
\affiliation{Center for Quantum Information Physics, New York University, New York, NY 10003, USA}
\author{S. M. Farzaneh}
\affiliation{Center for Quantum Information Physics, New York University, New York, NY 10003, USA}
\author{Jacob Issokson}
\affiliation{Center for Quantum Information Physics, New York University, New York, NY 10003, USA}
\author{Enrico Rossi}
\affiliation{Department of Physics, William \& Mary, Williamsburg, Virginia 23187, USA} 
\author{Javad Shabani}
\affiliation{Center for Quantum Information Physics, New York University, New York, NY 10003, USA}
\date{\today}

\begin{abstract}

Planar Josephson junctions (JJs) have emerged as a promising platform for the realization of topological superconductivity and Majorana zero modes. 
To obtain robust quasi one-dimensional (1D) topological superconducting states using 
planar JJs, limiting the number of 1D Andreev bound states' subbands that can be present, and increasing
the size of the topological superconducting gap, are two fundamental challenges. 
It has been suggested that both problems can be addressed
by properly designing the interfaces between the JJ's normal region and the superconducting leads.
We fabricated Josephson junctions with periodic hole structures on the superconducting contact leads on InAs heterostructures with epitaxial superconducting Al. By depleting the chemical potential inside the hole region with a top gate, we observed an enhancement of the supercurrent across the junction.
%
Such an enhancement is reproduced in theoretical simulations. The theoretical analysis shows that the enhancement of the JJ's critical current 
is achieved when the hole depletion is such to optimize the matching of quasiparticles' wave-function at the normal/superconductor interface.
These results show how the combination of carefully designed patterns for the Al coverage, and external gates, can be 
successfully used to tune the density and wave functions' profiles in the normal region of the JJ, and therefore
open a new avenue to tune some of the critical properties, such as number of subbands and size of the topological gap,
that must be optimized to obtain robust quasi 1D superconducting states supporting Majorana bound states.
%

\end{abstract}

\maketitle
Planar Josephson junctions based on 2DEGs with spin-orbit coupling and induced superconductivity have attracted a lot of attention in recent years, mainly due to their potential to realize topological superconductivity and Majorana zero modes (MZM) \cite{BG5,melo2019,BG3,BG4,BG6,BG1}. Under the presence of a suitable Zeeman field, the normal region in the Josephson junction can serve as a quasi-one-dimensional channel, which can host Majorana zero modes at its ends \cite{hell2017_planarJJ,pientka2017}. While solid progress has been made in both planar Josephson junctions and hybrid nanowire systems toward the realization of MZM \cite{mourik2012,deng2016,chen2017,shabani2016,ren2019,fornieri2019,dartiailh2021,qu2024}, so far an unambiguous demonstration of MZM is still missing \cite{Sarma2021,yu2021,pan2020,chen2019}. More and more evidence indicates that in hybrid systems there is a competition between disorder and the topological gap where disorder is still too strong \cite{ahn2021, Yu2023}. To make the hybrid system more robust against disorder, a large topological gap is preferable. As spin-orbit coupling and induced gaps jointly determine the size of the topological gap at finite magnetic fields \cite{stanescu2011}, an ideal hybrid system should have strong spin-orbit coupling and a large induced gap. Such improvements are feasible with the development of new combinations of semiconductors and superconductors. For example, the recent introduction of Sn and Pb into the nanowire system brings a much larger induced gap \cite{pendharkar2021, Kanne2021}. However, new materials also raise new challenges in material growth and device fabrication. InAs quantum well with epitaxial Aluminum, so far, is still the most suitable choice between nanofabrication and quality.

Compared to hybrid nanowire systems, planar Josephson junctions have a great advantage as the junction geometry 
can be easily modified to achieve strengthened properties. 
In particular, it might be possible to design the carrier density profiles and realize
``wave-function'' engineering using a combination of Al coverage patterns and external gates to minimize the number 
of subbands of Andreev bound states (ABSs),
and maximize the topologcal superconducting gap.
Several theoretical proposals also suggest modification of the Josephson junction geometry could lead to an enhanced induced gap and even an enhanced Rashba spin-orbit coupling \cite{laeven2020,paudel2021,melo2019}. 
With these potential benefits, it is natural to explore different geometries and utilize this degree of freedom to improve the system. 

\begin{figure}[h!]
\centering
  \includegraphics[width=\columnwidth]{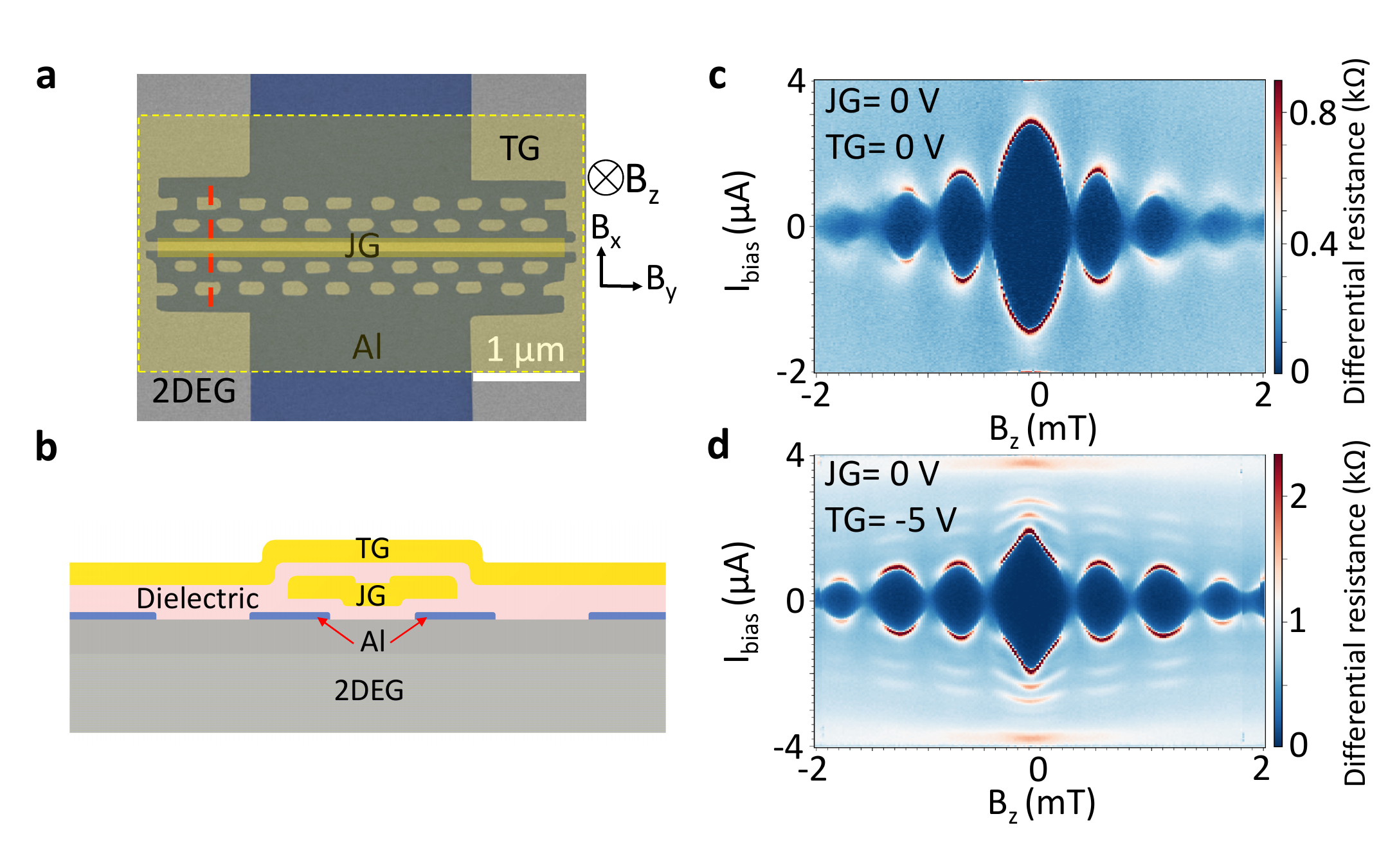}
  \caption{\textbf{Device geometry and Fraunhofer patterns at different gate configurations.} (a) False-color scanning electron micrograph of the measured device.  (b) schematic of the device and the material stacks. (c),(d) Differential resistance as a function of the bias current and out-of-plane magnetic field for JG = 0 V and TG = 0 V (c) and JG = 0 V and TG = -5 V  (d).
 }
 \label{fig1}
\end{figure}

We fabricated Josephson junctions on epitaxial superconducting Al thin films grown in-situ on InAs heterostructures. Such devices have shown high transparency 
\cite{mayer_superconducting_2019, kjaergaard_transparent_2017, dartiailh_missing_2021,elfeky2023} and a spin-orbit induced anomalous phase \cite{mayer_gate_2020}. The Josephson junction is 4 $\mu$m long with a width of around 100 nm. Two rows of periodic holes are etched on each side of the Al contacts together with the junction gap as shown in \cref{fig1}(a). Each hole is approximately 110 nm wide and 220 nm long. To control the chemical potential in the junction as well as in the hole region, we fabricated two layers of gates. In the first layer, a junction gate (JG) that covers the middle section of the junction is used to control the chemical potential in the junction. In the second layer, which is separated from the first layer of the gate by a second layer of dielectric, a top gate (TG) covers a much larger device region that includes the holes and the ends of the junction. A schematic diagram of the device and the material stacks are presented in \cref{fig1}(b). As shown in \cref{fig1}(a), the JG is shorter than the junction by 100 nm at each end by design. While JG itself cannot fully deplete the junction since it does not cover the ends of the junction, the chemical potential in the whole junction can still be fully controlled by using JG and TG together. We notice TG depletes the ends of the junction around -2.5 V (Shown in Fig.~S1 in Suppl Materials). Since the 2DEG inside the hole region should have a similar density and coupling to TG as the 2DEG in the junction, we expect that the 2DEG in the hole region should also be depleted around -2.5 V by TG. Tuning the chemical potential of the 2DEG in the hole region has a nontrivial effect on the junction, 
as discussed below. All the measurements in this study are performed in a dilution refrigerator equipped with a three-axis vector magnet. As shown in \cref{fig1}(a), the z-axis of the magnet is perpendicular to the device plane, while x and y-axes are in-plane fields aligned parallel and perpendicular to the current, respectively. Differential resistance is measured using standard low-frequency lock-in techniques in a four-point manner (more details about the device fabrication and measurement can be found in the methods section).
\begin{figure}[t!]
\centering
  \includegraphics[width=\columnwidth]{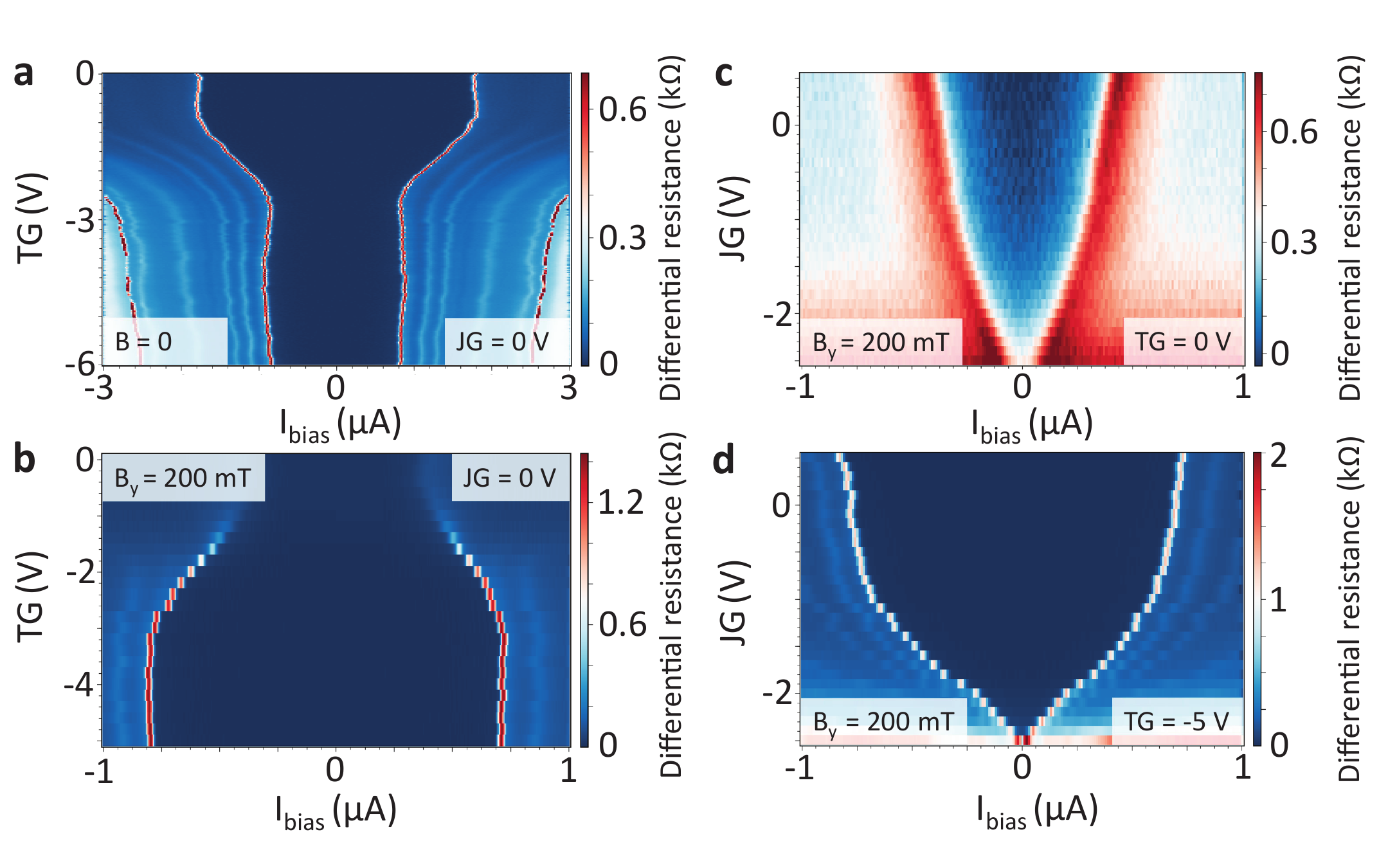}
  \caption{\textbf{Supercurrent gate dependence at different in-plane magnetic fields.} (a) Differential resistance as a function of the bias current and TG voltages at zero magnetic field. (b) Differential resistance as a function of the bias current and TG voltages at $B_y$ = 200 mT. The Supercurrent is significantly enhanced when a more negative voltage is applied to TG. (c),(d) Differential resistance as a function of the bias current and JG voltages at $B_y$ = 200 mT when TG = 0 V (c) and TG = - 5 V (d). Supercurrent is always monotonically decreasing with decreasing JG voltages.  }
  
 \label{fig2}
\end{figure}
\begin{figure}[h!]
\centering
  \includegraphics[width=\columnwidth]{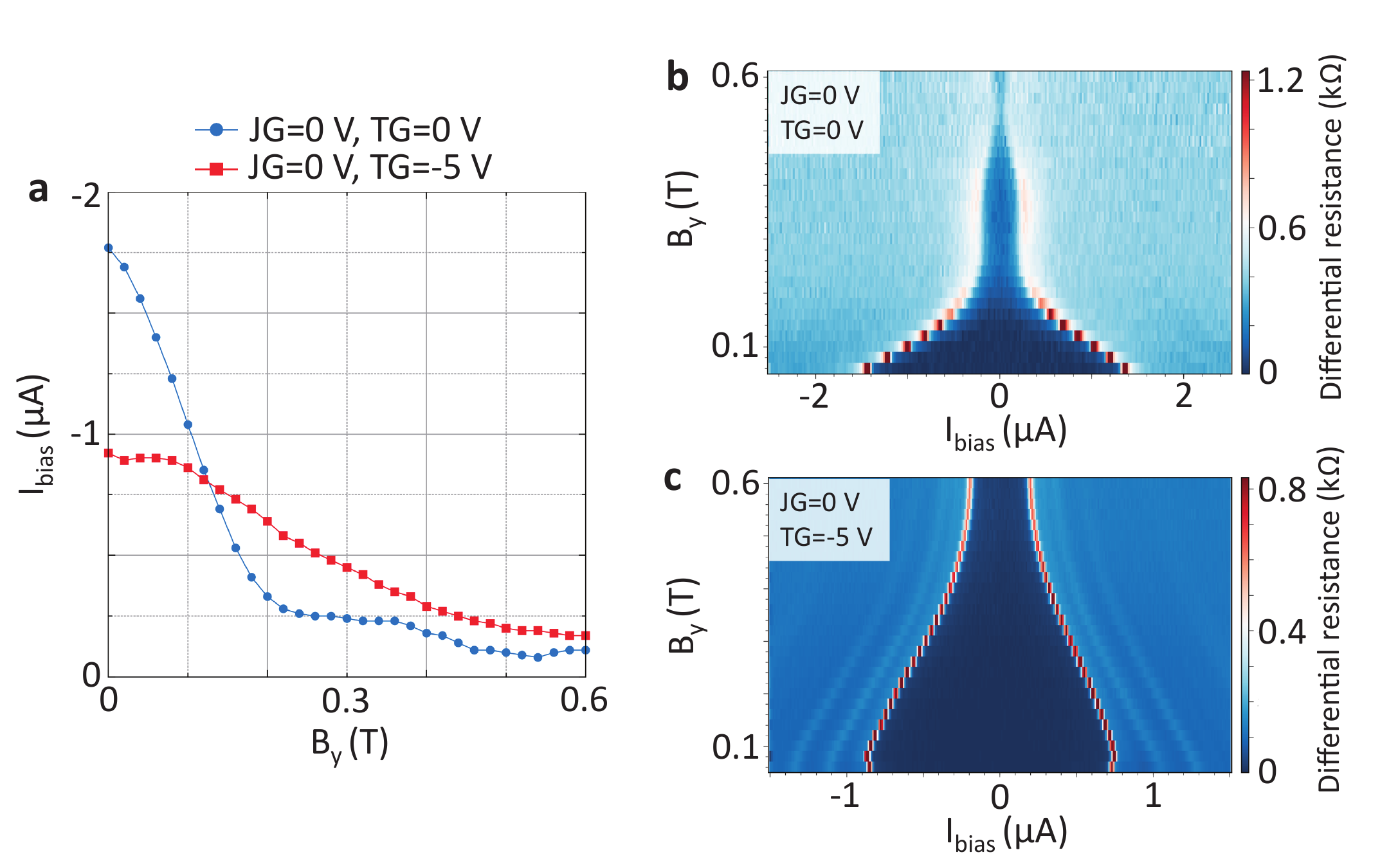}
  \caption{\textbf{Supercurrent in-plane field dependence at different gate configurations.} (a) Switching current extracted from panels (b) and (c). (b) Differential resistance as a function of the bias current and $B_y$ when JG = 0 V and TG = 0 V. (c) Differential resistance as a function of the bias current and $B_y$ when JG = 0 V and TG = -5 V, the supercurrent shows an almost linear decreasing with increasing $B_y$.  
 }
 \label{fig3}
\end{figure}
In \cref{fig1} (c)(d), we present the differential resistance as a function of the bias current and applied out-of-plane magnetic field for two different gate configurations. To eliminate the hysteresis due to heating effects, the current bias is swept from zero to high bias in these two scans. All gates are set to 0 for \cref{fig1}(c). For the results shown in \cref{fig1}(d), TG is set to -5 V while JG remains at 0. In \cref{fig1}(d), all the holes are supposed to be depleted as the result of a TG voltage below -2.5 V. In both configurations, the observed Fraunhofer patterns are symmetric with respect to the bias and the out-of-plane field, indicating an absence of hysteresis and a uniform supercurrent distribution across the junction. It is worth noting that extra resistance peaks have been observed in \cref{fig1}(d), which possibly indicates a stronger multiple Andreev reflection when the hole region is depleted. Overall, the two Fraunhofer patterns show similar periodicity, suggesting the effective junction area is not significantly modified by the depletion of the hole region.    
\begin{figure}[ht!]
    \centering
    \includegraphics[width=.48\textwidth]{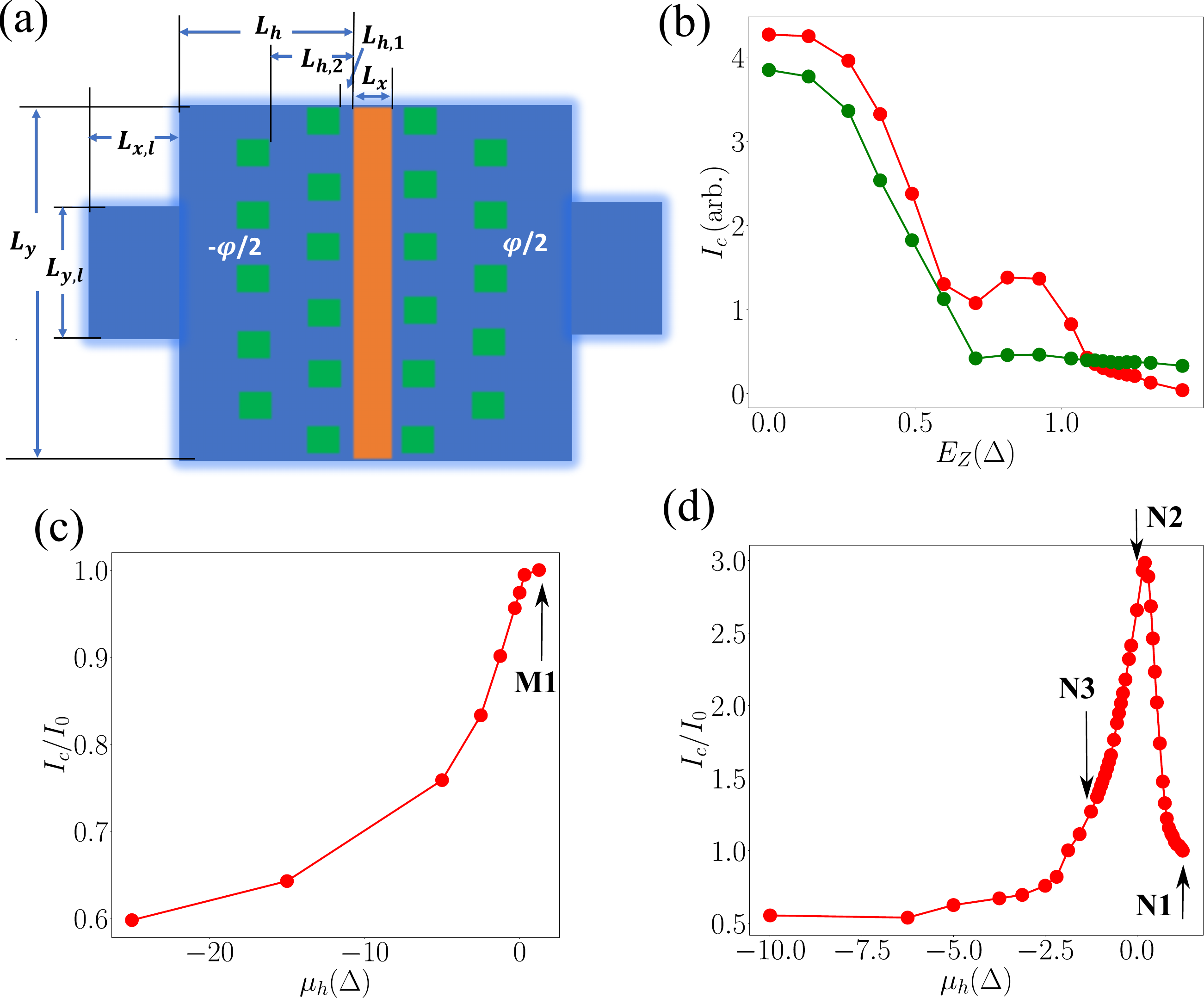}
    \caption{(a) Schematic of simulation setup to model the experiments. 
    (b) $I_c$ as a function of $E_Z$ for $\mu_h=\mu=1.25\Delta$ (red) and $\mu_h=-1.25\Delta$ (green).
    (c), (d) $I_c$ vs $\mu_h$ for $E_Z=0$ and $E_Z=1.14\Delta$, respectively.
    In (c) and (d) $I_0$ is the value of $I_c$ when $\mu_h=\mu$.
    %
    }
    \label{fig:sim}
\end{figure}
Next, we characterize the supercurrent gate dependence at different magnetic fields. At zero field, the supercurrent decreases when the voltage on TG is swept from 0 to -2.5 V as the result of the depletion of the ends of the junction (\cref{fig2}(a)). Below TG = -2.5 V, however, the supercurrent remains almost unchanged. This is consistent with the result in Fig.s1 which shows that TG depletes the ends of the junction around -2.5 V. 
For JG, the dependence is simpler as the supercurrent always monotonically decreases with decreasing JG voltages (Fig.s1(c)(d)). When the magnetic field is turned on, the supercurrent shows a very different behavior regarding TG voltages. At $B_y$ = 200 mT, where $B_y$ is perpendicular to the supercurrent, the supercurrent is significantly enhanced when a more negative voltage is applied to the TG (\cref{fig2}(b)). The supercurrent reaches its maximum around TG = -2.5V and remains almost constant for TG voltages below that value. As the depletion of the ends of the junction around TG = -2.5 V should only lead to a decrease of the supercurrent instead of an increase of the supercurrent, we suspect the abnormal enhancement of the supercurrent at finite fields could be related to the decreased chemical potential in the TG-controlled hole region. For fields parallel to the supercurrent, we observed a similar enhancement of supercurrent with TG voltages (Fig.s1(b)). When TG is fixed to 0 V and -5 V and $B_y$ = 200 mT, sweeping JG  reveals that the supercurrent still monotonically decreases with decreasing JG voltages (\cref{fig2}(c)(d)). That indicates the enhancement of supercurrent is solely determined by the change of the chemical potential in the hole region and is not related to the change of the chemical potential in the junction.\par 
The enhancement of the supercurrent can also be observed in in-plane magnetic field scans when the gate voltages are fixed. In \cref{fig3}(b), we plot differential resistance as a function of $B_y$ and bias current for TG = 0 V and JG = 0 V. The supercurrent exhibits a nonlinear behavior as it first quickly decreases at low fields followed by a much slower decline at higher fields. The supercurrent also shows some wiggles at higher fields. Overall, the supercurrent at $B_y$ = 0.5 T is less than 10 percent of the supercurrent at $B_y$ = 0.1 T. We notice the differential resistance of the "supercurrent" is finite at high fields, which is occasionally seen in our junctions. We attribute it to some part of the junction becoming normal at high fields. But the sharp drop in resistance and the Fraunhofer patterns at high fields (Fig.s2 ) indicate there is still supercurrent flow through the semiconductor part of the junction.  When TG is fixed to -5 V, \cref{fig3}(c), 
the supercurrent has an almost linear dependence on $B_y$. From $B_y$ = 0.1 T to $B_y$ = 0.5 T, the supercurrent still preserves 40 percent of its value and the resistance remains to zero. In \cref{fig3}(a), we extract the switching current from \cref{fig3}(b)(c) and plot them together. As can be seen, the two switching currents cross around $B_y$ = 0.12 T, confirming the observation that the enhancement of the supercurrent only happens at finite fields and when the voltage on TG is below the specific value.  In another device, we have observed a similar enhancement of supercurrent at finite fields when TG is below a certain value (see Supplemental Material (SM) for more details).    

\begin{figure}
    \centering
    \includegraphics[width=1.\linewidth]{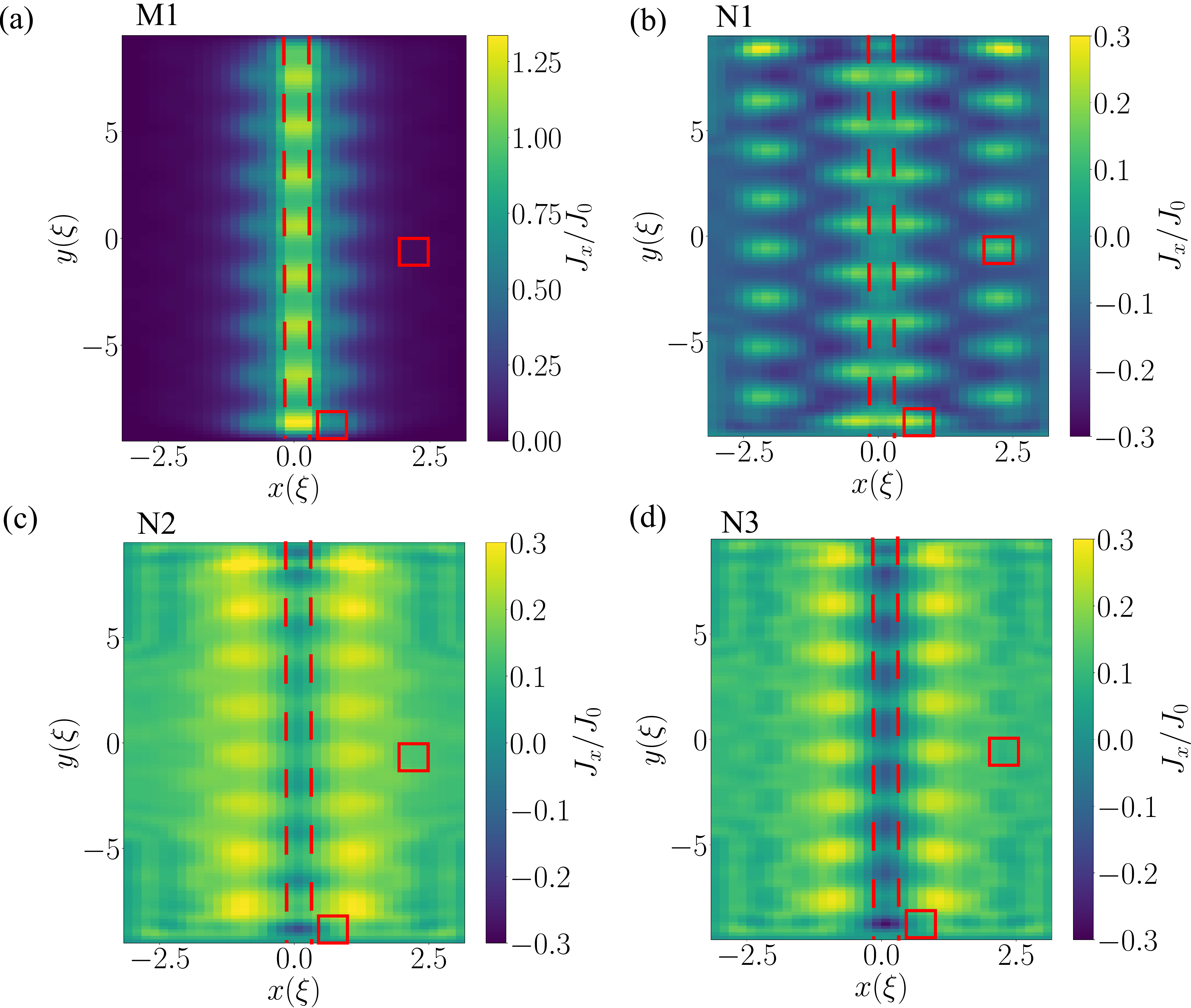}
    \caption{(a)
    Profile of $J_x$ for the case when $E_Z=0$ and $\mu_h=\mu$, M1 point in Fig.~\ref{fig:sim}~(c).
    (b), (c), (d) Profiles of $J_x$ for the case when $E_Z=1.14\Delta$ 
    and $\mu_h$ corresponds to the points N1, N2, N3 in Fig.~\ref{fig:sim}~(d), respectively.
    $J_0$ is the average current density for the case when $E_Z=0$, $\mu_h=\mu$.
    The red dashed lines indicate the boundary of the normal strip. The red boxes show positions of some of the depleted holes.
    }
    \label{fig:current1}
\end{figure}

\begin{figure}[ht!]
    \centering
    \includegraphics[width=.47\textwidth]{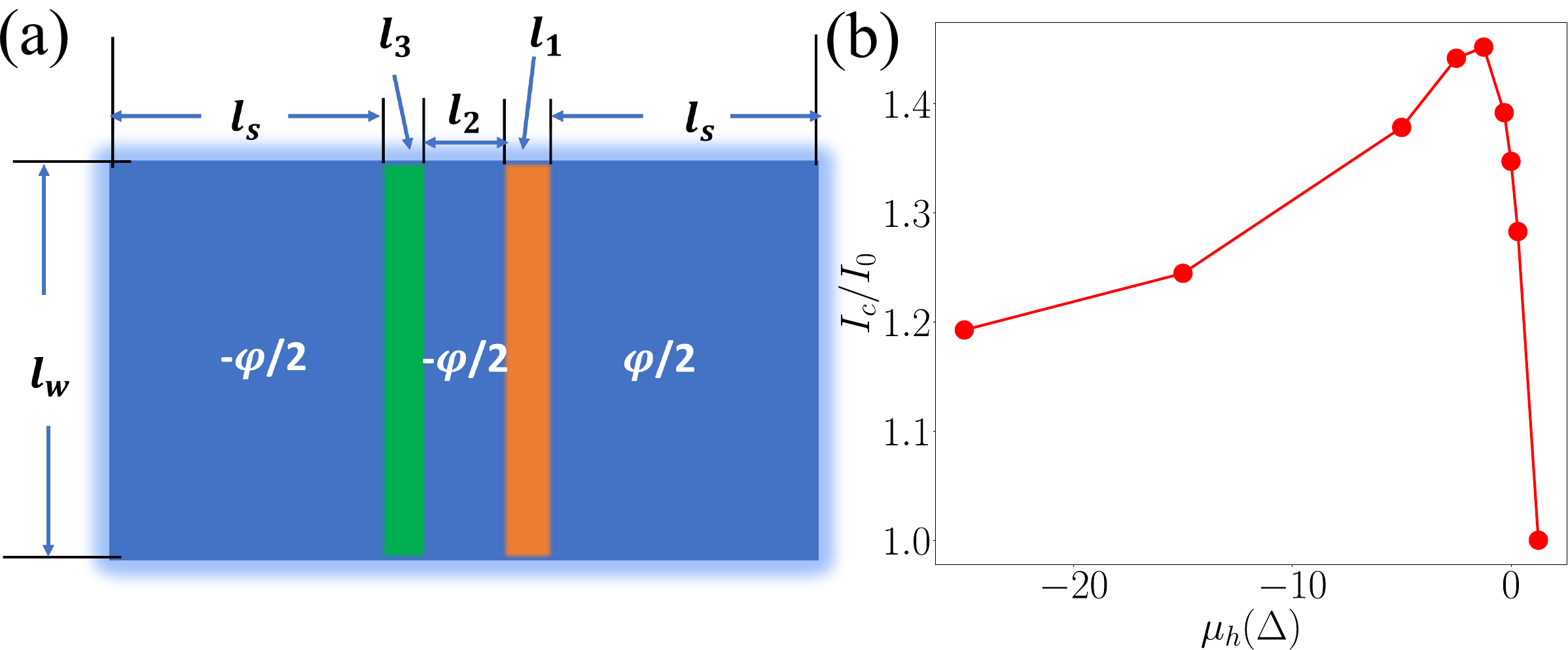}
    \caption{
    (a) Layout of simplified two-strip model.  
    (b) Scaling of $I_c$ with $\mu_h$ obtained using the simplified model for the case when $l_2=\xi/2$. $I_0=I_c(\mu_h=\mu)$.
    }
    %
    \label{fig:strip}
\end{figure}

To understand the origin of this dependence of critical currents on the gate voltage, 
we setup a tight-binding model for the Boguliobov de Gennes (BdG) Hamiltonian describing the system 
(see SM for details), using the python package Kwant~\cite{groth2014}.
To be able to obtain from the model all the desired quantities, in particular the critical current,  with the available computational resources,
we scaled down all the dimensions while using a value of the superconducting gap $\Delta$ and of the chemical potential
$\mu$ in the regions outside the holes such that
the ratio between the geometric dimensions and the superconducting coherence length, $\xi$,
is of the same order as in the experiment. For the results presented below we use a value of $\Delta$
16 times larger than the physical one and set $\mu=1.25\Delta$. The geometry of the model used is shown in Fig.~\ref{fig:sim}~(a).
To properly model the superconducting leads 
we chose a value of $L_{x,l}$,  see Fig.~\ref{fig:sim}~(a), 
sufficiently larger than $\xi$ 
to avoid spurious finite-size effects.

We first obtain the ABSs's spectrum $\{\epsilon_n(\varphi)\}$
as a function of the phase difference $\varphi$ 
between the superconducting pairing of the two leads. 
We then calculate the supercurrent $I(\varphi)=\sum_{\epsilon_n<0}(\partial{\epsilon_n}/{\partial\varphi})(2\pi/\Phi_0)$, 
where $\Phi_0=h/2e$ is the magnetic flux quantum.
From this, the critical current $I_c={\rm max}(I(\varphi))$ is extracted
for different values of the in-plane magnetic field and chemical potential $\mu_h$ of the holes.

Figure~\ref{fig:sim}~(b) shows the evolution of $I_c$ with the strenghth of the Zeeman energy $E_Z$
due to an in-plane magnetic field perpendicular to the current
for the case when $\mu_h=\mu$, in red, and $\mu_h=-1.25\Delta$, in green. The Zeeman energy in the superconductor is taken to be $1/2$ the value in the normal and depleted regions. 
In the first case the holes are normal regions, i.e., regions where $\Delta$ is set to zero,
with the same carrier density as the rest of the system. In the second case 
the carrier density in the holes is lower than in the areas around them.
In both cases we see that $I_c$ first decreases with $E_Z$ up to $E_Z\approx 0.6\Delta$.
For $E_z>0.6\Delta$ in the first case $I_c$ changes non-monotonically with $E_Z$,
a behavior that can be attributed to the almost closing and reopening of the gap of the ABS's spectrum
In the second case $I_c$ keeps decreasing also for $E_z>0.6\Delta$, albeit more slowly,
suggesting that in this case the gap of the ABS doesn't recover with $E_Z$.

As in the experimental case, from Fig.~\ref{fig:sim}~(b), we see that above a threshold value of $E_Z$, $E_Z\approx 1.1\Delta$,
$I_c$ for the depleted case is larger than for the non-depleted suggesting that
for $E_Z>1.1\Delta$ the evolution of $I_c$ with $\mu_h$ might be not monotonic.
This is confirmed by the results for the evolution of $I_c$ with respect to $\mu_h$
shown in Fig.~\ref{fig:sim}~(c),~(d), for the $E_Z=0$ and $E_Z=1.14\Delta$ cases, respectively.
We see that for $E_Z=0$, $I_c$ decreases monotonically as the depletion ($-\mu_h$) of the holes increases,
whereas for $E_Z=1.14\Delta$, $I_c$ varies non-monotonically with $\mu_h$, in qualitative agreement with 
the experimental results. 

We have verified, see SM, that different values of the spin-orbit coupling (SOC) strength,
and different directions of the in-plane magnetic field do not modify qualitatively
the results presented in Fig.~\ref{fig:sim}~(b)-(d).

Figure~\ref{fig:current1}
shows the spatial profile of the quasiparticle current density 
$J_x(x,y)=-i(\hbar/2m^*)\sum_{\epsilon_n<0}(\psi_n^*\nabla\psi_n-\psi_n\nabla\psi_n^*)$ \cite{BTK},
where $\psi_n$ are the eigenstates of the BdG Hamiltonian and $m^*$ is the effective mass. 
Notice that the regions where $J=0$ are regions where the current is carried by the superconducting condensate.
Panel (a) shows the profile of $J$ for the case when $E_Z=0$. 
We see that the presence of the holes induces a periodic modulation in the transverse direction
of the current in the normal region of the JJ.
Panels (b),~(c),~(d) show the results for the case $E_Z=1.14\Delta$ for
the values of $\mu_h$ denoted by $N1$, $N2$, and $N3$ in Fig.~\ref{fig:sim}~(d).
We see that for the value of $\mu_h$ for which $I_c$ is maximum, panel (c),
$J$ is more uniform in the central region of the JJ.
This suggest that the optimal value of $\mu_h$ results in a better matching
of the quasiparticle wave-functions across the different regions of the JJ.

To check that this is the case we considered a simplified model, shown in Fig.~\ref{fig:strip}~(a).
In this model the holes are effectively replaced by a
normal strip, shown in green in Fig.~\ref{fig:strip}~(a), at a distance $l_2$ from the normal region of the JJ.
Notice that the difference of the superconducting phase across the normal region modeling the holes
is set to zero. This is done to take into account that in the experimental geometry there are 
paths between the holes that connect the different superconducting regions 
on the same side of the JJ's normal region, the region shown in orange in Fig.~\ref{fig:strip}~(a),
that separates the left and right superconducting leads. 

Using the simplified model we were able to see that the critical parameter
determining the nature of the evolution of $I_c$ with respect to $\mu_h$
is the distance $l_2$ between the normal region modeling the holes, the depletion strip,
and the JJ normal region. 
Figure~\ref{fig:strip}(b) shows the result for $l_2=\xi/2$, when both $E_Z$ and SOC are not present. We find that when  $l_2\ll \xi$, $I_c$ decreases monotonically with ($-\mu_h$),
while at $l_2\sim\xi$, $I_c$ varies non-monotonically with (-$\mu_h$) as shown in Fig.~\ref{fig:strip}(b).
For $l_2\gg\xi$ the two normal strips are effectively decoupled from each other
and so $\mu_h$ has no effect on $I_c$.
These results suggest that the main reason why experimentally a non-monotonic scaling of $I_c$ vs (-$\mu_h$)
is observed in the presence of an in-plane magnetic field is the fact that when 
$E_Z\gtrsim\Delta$ the ABSs' wave functions decay faster in the superconducting regions
at the sides of the JJ's normal region, effect that in the simplified model corresponds to a 
reduction of $\xi$\footnote{In reality, the magnetic field will also suppress the superconducting gap, however, for the magnetic field considered, such effect is much smaller than the one arising from the reduction of the decay length of the ABS states.}
and therefore to an increase of the ratio $l_2/\xi$ to values $\sim1$ for which $I_c$ scales non-monotonically with $\mu_h$.

To further understand the origin of the observed dependence of $I_c$ on $l_2/\xi$ and $\mu_h$ we 
calculated the reflection coefficients of a superconducting-normal (SN) junction with a depletion region
on the superconducting side at a distance $l_2$ from the SN interface, see SM for details.
We found that the ratio $|r_A/r_N|$ between Andreev reflection, $r_A$, and normal reflection, $r_N$,
is strongly affected by $\mu_h$ when $l_2=\xi/2$ due to the fact that in this case
the amplitude of the electron wave function at the SN interface can be tuned in and out of the value that
maximizes $r_A$ by varying (-$\mu_h$). We therefore conclude that varying via an external gate 
the depletion of the holes for a JJ like the one shown in Fig.1~(a) results in the effective
tuning of the spatial profile of the quasiparticles wave functions leading to qualitative
changes of critical properties such as the JJ's critical current.

In summary, we have fabricated Josephson junctions with periodic hole structures on the Al contact. 
A counter intuitive enhancement of supercurrent has been observed when the 2DEG in the hole region is depleted by the TG. 
Theoretical modeling and careful analysis of the experimental results show that such unusual enhancement of
the critical current is due to changes of the spatial profile of the quasiparticles' wave functions.
The ability to shape engineer the wave function of the electronic quantum states is critical to
realize robust topological superconducting states supporting non-Abelian quasiparticles.
Our results show that by combining specific Al coverage layouts in InAs/Al planar JJs with 
external gates a unique control of the profile of the electrons' wave functions can be achieved
resulting in remarkable tunability of key properties of the JJs.



\subsection{Methods}

Wafers are grown by molecular beam epitaxy. Devices are fabricated using a combination of wet etching and deposition techniques after electron beam lithography. Device mesa features are defined by a deep wet etch with 85\% concentrated phosphoric acid, 30\% concentrated hydrogen peroxide, and deionized water in a volumetric ratio of 1:1:40 after selectively etching the aluminum top layer with Transene Aluminum Etchant Type D. Junction weak links and smaller device features are defined by a subsequent aluminum etch. Double-layer gates subsequently undergo two cycles of dielectric deposition of aluminum oxide via atomic layer deposition, and titanium/gold gates are deposited via electron beam evaporation. Measurements are performed in a dilution refrigerator at a temperature of around 30 mK using standard low-frequency lock-in amplification techniques with excitation currents of at most 10 nA and frequencies of around 17 and 77 Hz. Magnetic field is generated by a three-axis superconducting vector magnet.

\subsection{Author Contributions}
W.M.S. grew the material heterostructure. P.Y. fabricated the devices. P.Y. with the help of W.~F.~S., B.~H.~E, S.~M.~F. performed the measurements. H.F. and E.R. developed the theoretical description and obtained the theoretical results. P.Y., H.F., E.R. and J.S.
analyzed the results and wrote the manuscript with contributions from all of the authors.

\subsection{Acknowledgements}

Project was funded by the US Department of Energy, Office of Basic Energy Sciences, via Award DE-SC0022245. H.F. and E.R. thank Joseph J. Cuozzo for very helpful discussions during the initial stages of the work.

\bibliography{Ref.bib}
\end{document}